\documentclass[10pt]{article}
\usepackage{amssymb,amsmath,amsthm}
\usepackage{epsfig}
\newcommand{\ie}{\emph{i.e.}}
\newcommand{\eg}{\emph{e.g.}}
\newcommand{\cf}{\emph{cf}}

\newcommand{\etc}{\emph{etc}}

\newcommand{\Real}{\mathbb{R}}

\newcommand{\sii}{L^2}
\newcommand{\Sobi}{W^{1,2}}
\newcommand{\sobi}{W^{2,2}}

\newcommand{\Smooth}{C}
\newcommand{\Dom}{D}

\newcommand{\demi}{\frac{1}{2}}
\newcommand{\sgn}{\mathop{\mathrm{sgn}}\nolimits}
\newcommand{\diag}{\mathop{\mathrm{diag}}\nolimits}
\newcommand{\tubemap}{\mathcal{L}}
\newcommand{\tubemapalt}{\mathfrak{L}}
\newcommand{\Hilbert}{\mathcal{H}}
\newcommand{\Ball}{\mathcal{D}}
\newcommand{\Annulus}{\mathcal{A}}
\newcommand{\eps}{\varepsilon}
\newtheorem{Lemma}{Lemma}
\newtheorem{Proposition}{Proposition}
\newtheorem{Corollary}{Corollary}
\newtheorem{Theorem}{Theorem}
\theoremstyle{definition}
\newtheorem{Definition}{Definition}
\newtheorem{Remark}{Remark}
\begin{document}
%
\title{\textbf{\Large
Waveguides with combined
Dirichlet and Robin boundary conditions
}}
\author{\textsc{
P.~Freitas$^1$
\ and \
D.~Krej\v{c}i\v{r}\'{\i}k$^2$
}}
\date{
\footnotesize
\begin{quote}
\emph{
\begin{itemize}
\item[$1$]
Department of Mathematics, Faculdade de Motricidade Humana (TU Lisbon)
{\rm and} Group of Mathematical Physics of the University of Lisbon,
Complexo Interdisciplinar, Av.~Prof.~Gama Pinto~2,
P-1649-003 Lisboa, Portugal
\smallskip \\
\emph{E-mail:} freitas@cii.fc.ul.pt
\smallskip
\item[$2$]
Department of Theoretical Physics, Nuclear Physics Institute, \\
Academy of Sciences,
250\,68 \v{R}e\v{z} near Prague, Czech Republic
\smallskip \\
\emph{E-mail:} krejcirik@ujf.cas.cz
\end{itemize}
}
\end{quote}
}
\maketitle
%
\begin{abstract}
\noindent
We consider the Laplacian in a curved two-dimensional strip
of constant width squeezed between two curves,
subject to Dirichlet boundary conditions on one of the curves
and variable Robin boundary conditions on the other.
We prove that,
for certain types of Robin boundary conditions,
the spectral threshold of the Laplacian is estimated
from below by the lowest eigenvalue of the Laplacian in
a Dirichlet-Robin annulus determined by the geometry of the strip.
Moreover, we show that an appropriate combination of the geometric setting
and boundary conditions leads to a Hardy-type inequality
in infinite strips.
As an application, we derive certain stability of the spectrum
for the Laplacian in Dirichlet-Neumann strips
along a class of curves of sign-changing curvature,
improving in this way an initial result
of Dittrich and K\v{r}{\'\i}\v{z}~\cite{DKriz2}.
\medskip
\begin{itemize}
\item[\textbf{MSC\,2000:}]
35P15; 58J50; 81Q10.
\item[\textbf{Keywords:}]
Dirichlet and Robin boundary conditions;
Eigenvalues in strips and annuli;
Hardy inequality;
Laplacian;
Waveguides.
\end{itemize}
\end{abstract}
%

\newpage
%
\section{Introduction}\label{Sec.Intro}
%
The Laplacian in an unbounded tubular region~$\Omega$
has been extensively studied as a reasonable model for the Hamiltonian
of electronic transport in long and thin semiconductor structures
called \emph{quantum wave\-guides}.
We refer to~\cite{DE,LCM} for the physical background and references.
In this model, it is more natural to consider
Dirichlet boundary conditions on~$\partial\Omega$
corresponding to a large chemical potential barrier (\cf~\cite{ES,GJ,DE}).

However, Neumann boundary conditions
or a combination of Dirichlet and Neumann boundary conditions
have been also investigated.
We refer to~\cite{KZ1,KZ2} for the former
and to~\cite{DKriz2,OM,KKriz} for the latter.
Moreover, these types of boundary conditions
are relevant to other physical systems
(\cf~\cite{ELV,DP,JLP}).

Although we are not aware of any work in the literature where more
general boundary conditions have been considered in the case of quantum waveguides,
it is possible to think also of Robin boundary conditions as
modelling impenetrable walls of~$\Omega$ in the sense that there is no
probability current through the boundary.
Furthermore, Robin boundary conditions may in principle
be relevant for different types of inter\-phase in a solid.

Moreover, the interplay between boundary conditions, geometry
and spectral properties is an interesting mathematical problem in itself.
To illustrate this, let us recall that
it has been known for more than a decade
that the curved geometry of an unbounded planar strip
of uniform width may produce eigenvalues below the essential spectrum.
We refer to the pioneering work~\cite{ES} of Exner and \v{S}eba
and the sequence of papers~\cite{GJ,RB,DE,KKriz,ChDFK}
for the existence results under rather simple and general
geometric conditions.

However, it has not been noticed until the recent letter~\cite{DKriz2}
of Dittrich and K\v{r}{\'\i}\v{z}
that the existence of eigenvalues in fact
depends heavily on the geometrical setting
provided the uniform Dirichlet boundary conditions
are replaced by a combination of Dirichlet and Neumann ones.
In particular, the discrete spectrum may be eliminated
provided the Dirichlet-Neumann strip is ``curved appropriately'',
\ie, the Neumann boundary condition is imposed
on the ``locally shorter'' boundary curve.

Recently, it has also been shown that the discrete spectrum
may be eliminated by adding a local magnetic field
perpendicular to a planar Dirichlet strip \cite{MK-Kov,B-MK-Kov},
by embedding the strip into a curved surface \cite{K3}
or by twisting a three-dimensional Dirichlet tube
of non-circular cross-section \cite{EKK}.

The aim of the present paper is to examine further
the interplay between boundary conditions, geometry
and spectral properties in the case of~$\Omega$
being a planar strip with a combination of Dirichlet
and (variable) Robin boundary conditions on~$\partial\Omega$.
Our main result is a lower bound to the spectral threshold
of the Laplacian in
a (bounded or unbounded) Dirichlet-Robin strip.
This enables us to prove quite easily non-existence results
about the discrete spectrum for certain waveguides,
and generalize in this way the results
of Dittrich and K\v{r}{\'\i}\v{z}~\cite{DKriz2}.
Moreover, we show that certain combinations
of boundary conditions and geometry
lead to Hardy-type inequalities
for the Laplacian in unbounded strips.
These inequalities are new in the theory of quantum waveguides
with combined boundary conditions.
As an application, we further extend the class of
Dirichlet-Neumann strips with empty discrete spectrum.

\section{Scope of the paper}
%
In this section we precise the problem
we deal with in the present paper
and state our main results.

\subsection{The model}
%
Given an open interval $I\subseteq\Real$
(bounded or unbounded),
let
$
  \Gamma\equiv(\Gamma^1,\Gamma^2):I\to\Real^2
$
be a unit-speed $\Smooth^2$-smooth plane curve.
We assume that~$\Gamma$ is an embedding.
The function $N:=(-\dot{\Gamma}^2,\dot{\Gamma}^1)$ defines
a unit normal vector field along~$\Gamma$
and the couple $(\dot{\Gamma},N)$
gives a distinguished Frenet frame
(\cf~\cite[Chap.~1]{Kli}).
The curvature of~$\Gamma$
is defined through the Serret-Frenet formulae by
$\kappa:=\det(\dot{\Gamma},\ddot{\Gamma})$;
it is a continuous function of the arc-length parameter.
We assume that~$\kappa$ is bounded.
It is worth noticing that the curve~$\Gamma$ is fully determined
(except for its position and orientation in the plane)
by the curvature function~$\kappa$ alone (\cf~\cite[Sec.~II.20]{Kreyszig}).

Let~$a$ be a given positive number.
We define the mapping
\begin{equation}\label{StripMap}
  \tubemap: I \times [-a,a] \to \Real^2:
  \big\{ (s,t) \mapsto \Gamma(s)+ N(s) \, t \big\}
\end{equation}
and make the hypotheses that
\begin{equation}\label{Ass.Basic}
  \|\kappa\|_\infty \, a < 1
  \qquad\mbox{and}\qquad
  \tubemap \quad \mbox{is injective}.
\end{equation}
Then the image
\begin{equation}\label{image}
  \Omega := \tubemap\big(I \times (-a,a)\big)
\end{equation}
has a geometrical meaning
of an open non-self-intersecting strip,
contained between the parallel curves
$$
  \Gamma_\pm:=\tubemap(I\times\{\pm a\})
$$
at the distance~$a$ from~$\Gamma$,
and, if~$\partial I$ is not empty, the straight lines
$L_-:=\tubemap\big(\{\inf I\}\times(-a,a)\big)$
and $L_+:=\tubemap\big(\{\sup I\}\times(-a,a)\big)$.
The geometry is set in such a way that~$\kappa>0$
implies that the parallel curve~$\Gamma_+$
is locally shorter than~$\Gamma_-$, and vice versa.
We refer to~\cite[App.~A]{EKK}
for a sufficient condition ensuring
the validity of the second hypothesis in~(\ref{Ass.Basic}).

Given a bounded continuous function $\tilde{\alpha}:\Gamma_+\to\Real$,
let $-\Delta_{\kappa,\alpha}$ denote
the (non-negative) Laplacian on $\sii(\Omega)$,
subject to uniform Dirichlet boundary conditions
on the parallel curve~$\Gamma_-$,
uniform Neumann boundary conditions on $L_- \cup L_+$
(\ie\ none if~$\partial I$ is empty)
and the Robin boundary conditions of the form
\begin{equation}\label{Robin}
  \frac{\partial u}{\partial N} + \tilde{\alpha} \,u = 0
  \qquad\mbox{on}\qquad \Gamma_+
  \,,
\end{equation}
where $u\in\Dom(-\Delta_{\kappa,\alpha})$.
Hereafter we shall rather use
$\alpha:=\tilde{\alpha}\big(\tubemap(\cdot,a)\big)$,
a function on~$I$.
Notice that the choice $\alpha=0$
corresponds to uniform Neumann boundary conditions on~$\Gamma_+$
and $\alpha\to+\infty$ approaches
uniform Dirichlet boundary conditions on~$\Gamma_+$;
for this reason, we shall sometimes use ``$\alpha=+\infty$"
to refer to the latter.
The Laplacian $-\Delta_{\kappa,\alpha}$ is properly defined
in Section~\ref{Sec.Laplacian} below
by means of a quadratic-form approach.

\subsection{A lower bound to the spectral threshold}
%
If the curvature~$\kappa$ is a constant function,
then the image~$\Omega$ can be identified
with a segment of an annulus or a straight strip.
We prove that, in certain situations,
this constant geometry minimizes the spectrum of $-\Delta_{\kappa,\alpha}$,
within all admissible functions~$\kappa$ and~$\alpha$
considered as parameters.

More precisely, let us denote by~$\Ball(r)$
the open disc of radius~$r>0$
and let
$
  \Annulus(r_1,r_2) :=
  \Ball(r_2)\setminus\overline{\Ball(r_1)}
$
be an annulus of radii $r_2>r_1>0$.
Abusing the notation for~$\kappa$ and~$\alpha$ slightly,
we introduce a function
$
  \lambda: (-a,a)\times\Real \to \Real
$
by means of the following definition:
\begin{Definition}\label{Def.lambda}
Given two real numbers~$\alpha$ and $\kappa$, with $\kappa$ in $(-1/a,1/a)$,
we denote by $\lambda(\kappa,\alpha)$
the spectral threshold of the Laplacian on
\begin{equation*}
  \Annulus_{\kappa} :=
\begin{cases}
  \Annulus\big(|\kappa|^{-1}-a,|\kappa|^{-1}+a\big)
  & \mbox{if} \quad \kappa\not=0 \,,
  \\
  \Real\times(-a,a)
  & \mbox{if} \quad \kappa=0 \,,
\end{cases}
\end{equation*}
subject to uniform Dirichlet boundary condition on
\begin{equation*}
\begin{cases}
  \partial\Ball(\kappa^{-1}+a)
  & \mbox{if} \quad \kappa>0 \,,
  \\
  \Real\times\{-a\}
  & \mbox{if} \quad \kappa=0 \,,
  \\
  \partial\Ball(|\kappa|^{-1}-a)
  & \mbox{if} \quad \kappa<0 \,,
\end{cases}
\end{equation*}
and uniform Robin boundary conditions of the type~(\ref{Robin})
(with~$\alpha$ constant
and~$N$ being the outward unit normal
on~$\partial\Annulus_{\kappa}$)
on the other connected part of the boundary.
\end{Definition}

The most general result of the present paper
reads as follows:
\begin{Theorem}\label{Thm.bound}
Given a positive number~$a$
and a bounded continuous function~$\kappa$,
let~$\Omega$ be the strip defined by~(\ref{image})
with~(\ref{StripMap}) and satisfying~(\ref{Ass.Basic}).
Let~$\alpha$ be a bounded continuous function.
Then
\begin{equation}\label{bound}
  \inf\sigma(-\Delta_{\kappa,\alpha})
  \geq \lambda(\inf\kappa,\inf\alpha)
  \quad
  \mbox{provided $\kappa \leq 0$ or $\alpha \leq 0$.}
\end{equation}
\end{Theorem}

\begin{figure}[t]
\epsfig{file=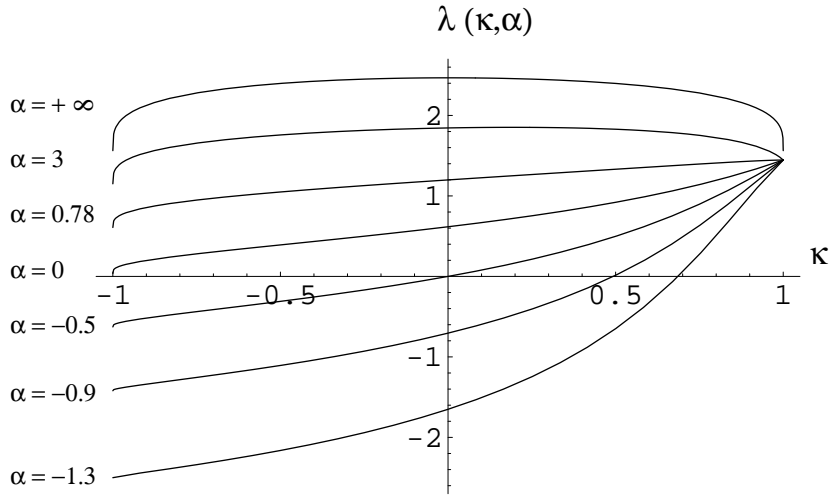,width=0.9\textwidth}
\caption{
Dependence of the lower bound~$\lambda(\kappa,\alpha)$
on the curvature~$\kappa$ for $a=1$ and different values of~$\alpha$.
(All curves meet at $\kappa=1/a$, the small gap for the curve
with~$\alpha=+\infty$ is due to a numerical inaccuracy.)
}
\label{figure}
\end{figure}

The lower bound~$\lambda(\kappa,\alpha)$ as a function of curvature~$\kappa$
for certain values of~$\alpha$ is depicted in Figure~\ref{figure}.
We prove the following properties which are important for~(\ref{bound}):
\begin{Theorem}\label{thm.annulus}
$\lambda$ satisfies the following properties:
\begin{itemize}
\item[\emph{(i)}]
$
  \forall\kappa\in(-1/a,1/a), \quad
  \alpha\mapsto\lambda(\kappa,\alpha): \Real\to\Real
$
is continuous and increasing,
\item[\emph{(ii)}]
$
  \forall\alpha\in\Real, \quad
  \kappa\mapsto\lambda(\kappa,\alpha): (-1/a,1/a) \to \Real
$
is continuous,
\item[\emph{(iii)}]
$
  \forall\alpha\in\Real, \quad
  \kappa\mapsto\lambda(\kappa,\alpha): (-1/a,0] \to \Real
$
is increasing,
\item[\emph{(iv)}]
$
  \forall\alpha\in(-\infty,0], \quad
  \kappa\mapsto\lambda(\kappa,\alpha): (-1/a,1/a) \to \Real
$
is increasing,
\item[\emph{(v)}]
$
  {\displaystyle
  \forall\alpha\in\Real, \quad
  \lim_{\kappa\to -1/a}\lambda(\kappa,\alpha)
  = \nu(\alpha) ,
  \ \,
  \lim_{\kappa\to 1/a}\lambda(\kappa,\alpha)
  = \nu(+\infty) ,
  }
$
\end{itemize}
where $\nu(\alpha)$, with $\alpha\in\Real\cup\{+\infty\}$,
denotes the first eigenvalue of the Laplacian in the disc $\Ball(2a)$,
subject to uniform Robin boundary conditions of the type~(\ref{Robin})
if $\alpha\in\Real$ (with $\alpha$ constant
and~$N$ being the outward unit normal on~$\partial\Ball(2a)$)
or uniform Dirichlet boundary conditions if $\alpha=+\infty$.
\end{Theorem}

Of course, $\nu(+\infty)=j_{0,1}^2/(2a)^2$,
where~$j_{0,1}$ denotes the first zero of the Bessel function~$J_0$,
and $\nu(0)=0$.

Theorem~\ref{Thm.bound} is a natural continuation of efforts
to estimate the spectral threshold in curved Dirichlet tubes~\cite{AE,EFK}.
More specifically, in the recent article~\cite{EFK},
Exner and the present authors establihed a lower bound of the type~(\ref{bound})
for the case~$\alpha=+\infty$, \ie~for pure Dirichlet strips
(the results in that paper are more general in the sense
that the tubes considered there
were multi-dimensional and of arbitrary  cross-section).
Namely,
$$
  \inf\sigma(-\Delta_{\kappa,+\infty})
  \geq \lambda\big(\|\kappa\|_\infty,+\infty\big)
  \,,
$$
where~$\lambda(\kappa,+\infty)$ is the spectral threshold
of the Dirichlet Laplacian in~$\Annulus_\kappa$.
It is also established in~\cite{EFK}
that $\kappa\mapsto\lambda(\kappa,+\infty)$
is an even function, decreasing on $[0,1/a)$,
reaching its maximum~$\pi^2/(2a)^2$ for $\kappa=0$ (a straight strip)
and approaching its infimum~$\nu(+\infty)$ as $\kappa\to 1/a$ (a disc).
The style and the main idea
(\ie~the intermediate lower bound~(\ref{inter.bound}) below)
of the present paper are similar to that of~\cite{EFK}. However,
we have to use different techniques to establish the
properties of~$\lambda$ (Theorem~\ref{thm.annulus}),
and consequently~(\ref{bound}).

\subsection{A Hardy inequality in infinite strips}
%
Theorem~\ref{Thm.bound} is optimal in the sense
that the lower bound~(\ref{bound}) is achieved
by a strip (along a curve of constant curvature).
On the other hand, since the minimizer is bounded
if the curvature is non-trivial,
a better lower bound is expected to hold for unbounded strips.
Indeed, in certain unbounded situations,
we prove that the lower bound of Theorem~\ref{Thm.bound}
can be improved by a Hardy-type inequality.

Let us therefore consider the infinite case $I=\Real$
in this subsection.
Let~$\alpha_0$ be a given real number.
If~$\kappa$ is equal to zero identically
(\ie~$\Omega$ is a straight strip)
and~$\alpha$ is equal to~$\alpha_0$ identically,
it is easy to see that
\begin{equation}\label{Spec.Straight}
  \sigma(-\Delta_{0,\alpha_0})
  = \sigma_\mathrm{ess}(-\Delta_{0,\alpha_0})
  = \big[\lambda(0,\alpha_0),\infty\big)
  \,.
\end{equation}

Although the results below hold
under more general conditions about vanishing of~$\kappa$
and the difference $\alpha-\alpha_0$ at infinity
(\cf~Section~\ref{Sec.closing} below),
for simplicity, we restrict ourselves to strips
which are deformed only locally in the sense
that~$\kappa$ and $\alpha-\alpha_0$
have compact support.
Under these hypotheses,
it is easy to verify that
the essential spectrum is preserved:
\begin{equation}\label{EssSpec}
  \sigma_\mathrm{ess}(-\Delta_{\kappa,\alpha})
  = \big[\lambda(0,\alpha_0),\infty\big)
  \,.
\end{equation}
A harder problem is to decide whether this interval
exhausts the spectrum of $-\Delta_{\kappa,\alpha}$,
or whether there exists discrete eigenvalues
below $\lambda(0,\alpha_0)$.

On the one hand,
Dittrich and K\v{r}{\'\i}\v{z}~\cite{DKriz2}
showed that the curvature
which is negative in a suitable sense
creates eigenvalues
below the threshold $\lambda(0,0)$
in the uniform Dirichlet-Neumann case
(\ie~in the case $\alpha=0$ identically).
For instance, using, in analogy to~\cite{DKriz2},
a modification of the ``generalized eigenfunction''
of~$-\Delta_{0,\alpha_0}$ corresponding to $\lambda(0,\alpha_0)$
as a test function, it is straightforward
to extend a result of~\cite{DKriz2}
to the case of uniform Robin boundary conditions:
\begin{Proposition}\label{Prop.DK}
Let $I=\Real$.
If $\alpha(s)=\alpha_0$ for all $s\in\Real$
and $\int_\Real \kappa(s) \, ds < 0$,
then
$$
  \inf\sigma(-\Delta_{\kappa,\alpha_0})
  < \lambda(0,\alpha_0)
  \,.
$$
\end{Proposition}
\noindent
In particular, Proposition~\ref{Prop.DK}
together with~(\ref{EssSpec}) implies
that the discrete spectrum of $-\Delta_{\kappa,\alpha_0}$ exists
if the strip is appropriately curved and asymptotically straight.
Notice also that the discrete spectrum
may be created by variable~$\alpha$
even if~$\Omega$ is straight
(\cf~\cite{EK1,EK2} for this type of results
in a similar model).

On the other hand,
Dittrich and K\v{r}{\'\i}\v{z}~\cite{DKriz2} showed
that the spectrum of~$-\Delta_{\kappa,0}$
coincides with the interval~(\ref{EssSpec}) with $\alpha_0=0$
provided the curvature~$\kappa$ is non-negative
and of compact support.
More precisely, they proved that
\begin{equation}\label{DKriz}
   \inf\sigma(-\Delta_{\kappa,0})
   \geq \lambda(0,0)
   \qquad\mbox{provided}\qquad
   \kappa \geq 0
   \,,
\end{equation}
which implies the result in view of~(\ref{EssSpec}).
Of course, not only the lower bound~(\ref{DKriz})
is contained in our Theorem~\ref{Thm.bound},
but the latter also generalizes the former
to variable Robin boundary conditions:
\begin{Corollary}\label{Corol.Infinite}
Let $I=\Real$ and assume that $\kappa$ and $\alpha-\alpha_0$
have compact support.
Under the hypotheses of Theorem~\ref{Thm.bound},
$$
  \sigma(-\Delta_{\kappa,\alpha})
  = \sigma_\mathrm{ess}(-\Delta_{\kappa,\alpha})
  = \big[\lambda(0,\alpha_0),\infty\big)
  \qquad\mbox{if}\qquad
  \kappa \geq 0 \,, \quad
  \alpha_0 \leq \alpha \leq 0
  \,.
$$
\end{Corollary}

Apart from this generalization,
Theorem~\ref{Thm.bound} provides an alternative
and, we believe, more elegant, proof of~(\ref{DKriz}).
Indeed, the proof of Dittrich and K\v{r}{\'\i}\v{z} in~\cite{DKriz2}
is very technical, based on a decomposition of $-\Delta_{\kappa,0}$
into an orthonormal basis and an analysis of solutions of Bessel type
to an associated ordinary differential operator,
while the proof of Theorem~\ref{Thm.bound}
does not require any explicit solutions whatsoever.

Furthermore, we obtain a stronger result,
namely, that a Hardy-type inequality
actually holds true in positively curved Dirichlet-Robin strips:
\begin{Theorem}\label{Thm.Hardy}
Let $I=\Real$.
Given a positive number~$a$
and a bounded continuous function~$\kappa$,
let~$\Omega$ be the strip defined by~(\ref{image})
with~(\ref{StripMap}) and satisfying~(\ref{Ass.Basic}).
Let~$\alpha$ be a bounded continuous function
such that $\alpha_0 \leq \alpha \leq 0$.
Assume that~$\kappa$ is non-negative
and that either one of~$\kappa$
or $\alpha-\alpha_0$ is not identically equal to zero.
Then, for any~$s_0$ such that $\kappa(s_0) > 0$
or $\alpha(s_0) > \alpha_0$, we have
\begin{equation}\label{Hardy.bound}
  -\Delta_{\kappa,\alpha}
  \ \geq \
  \lambda(0,\alpha_0)
  + \frac{c}{(\rho\circ\tubemap^{-1})^2}
\end{equation}
in the sense of quadratic forms
(\cf~(\ref{Hardy.bound.exact}) below).
Here~$c$ is a positive constant which depends
on~$s_0$, $a$, $\kappa$ and $\alpha$,
$
  \rho(s,t) := \sqrt{1+(s-s_0)^2}
$
and~$\tubemap$ is given by~(\ref{StripMap}).
\end{Theorem}

It is possible to find an explicit lower bound
for the constant~$c$;
we give an estimate in~(\ref{constant}) below.

Theorem~\ref{Thm.Hardy} implies
that the presence of a positive curvature
or of suitable Robin boundary conditions
represents a repulsive interaction in the sense that there
is no spectrum below~$\lambda(0,\alpha_0)$
for all small potential-type perturbations
having a sufficiently fast decay at infinity.
This provides certain stability of the spectrum
of the type established in Corollary~\ref{Corol.Infinite}.

Moreover, in the uniform Dirichlet-Neumann case,
we use Theorem~\ref{Thm.Hardy} to show
that the spectrum is stable even if~$\kappa$
is allowed to be negative:
\begin{Corollary}\label{Thm.stability}
Given a positive number~$a$
and a bounded continuous function~$\kappa$
of compact support,
let~$\Omega$ be the strip defined by~(\ref{image})
with~(\ref{StripMap}) and satisfying~(\ref{Ass.Basic}).
Assume that
$$
  |\kappa_-| \leq \eps
  \qquad\mbox{with}\qquad
  \eps \geq 0
  \,,
$$
while~$\kappa_+$ is independent of~$\eps$
and not identically equal to zero.
Then there exists a positive number~$\eps_0$
such that for all $\eps\leq\eps_0$ we have
\begin{equation*}
  \sigma(-\Delta_{\kappa,0})
  = \sigma_\mathrm{ess}(-\Delta_{\kappa,0})
  = \big[\lambda(0,0),\infty\big)
  \,.
\end{equation*}
Here~$\eps_0$ depends on~$a$, $\kappa_+$ and~$I$.
\end{Corollary}

Corollary~\ref{Thm.stability} follows
as a consequence of~(\ref{EssSpec})
and the Hardy inequality~(\ref{Hardy.general}) below.
This generalizes a result of~\cite{DKriz2}
to strips with sign-changing curvature.

\subsection{Contents}
%
The present paper is organized as follows.

In the following section
we introduce the Laplacian~$-\Delta_{\kappa,\alpha}$
in the curved strip $\Omega$ by means of
its associated quadratic form
and express it in curvilinear coordinates
defined by~(\ref{StripMap}).
We obtain in this way an operator
of the Laplace-Beltrami form
in the straight strip $I\times(-a,a)$.

In Section~\ref{Sec.Inter}
we show that the structure of the Laplace-Beltrami operator
leads quite easily to a ``variable'' lower bound~(\ref{inter.bound}),
expressed in terms of the function~$\lambda$ of Definition~\ref{Def.lambda}.
We call this lower bound ``intermediate''
since this and Theorem~\ref{thm.annulus}
imply Theorem~\ref{Thm.bound} at once.

In Section~\ref{Sec.Annulus}
we prove Theorem~\ref{thm.annulus} using a combination of a number of techniques,
such as the minimax principle, the maximum principle,
perturbation theory, \etc.

Section~\ref{Sec.Infinite} is devoted to infinite strips,
namely, to the proofs of Theorem~\ref{Thm.Hardy}
and its Corollary~\ref{Thm.stability}.
The former is based on an improved intermediate lower bound,
Theorem~\ref{thm.annulus} and the classical one-dimensional
Hardy inequality.

In the closing section
we discuss possible extensions
and refer to some open problems.

\section{The Laplacian}\label{Sec.Laplacian}
%
The Laplacian~$-\Delta_{\kappa,\alpha}$
is properly defined as follows.
We introduce on the Hilbert space $\sii(\Omega)$
the quadratic form~$Q_{\kappa,\alpha}$ defined by
\begin{equation}\label{form.Laplacian}
\begin{aligned}
  Q_{\kappa,\alpha}[u] &:= \int_{\Omega} |\nabla u(x)|^2 \, dx
  + \int_{\Gamma_+} \tilde{\alpha}(\sigma) \, |u(\sigma)|^2 \, d\sigma
  \,,
  \\
  u \in \Dom(Q_{\kappa,\alpha}) &:=
  \big\{
  u \in \Sobi(\Omega) \, \big| \
  u(\sigma)=0 \quad \mbox{for a.e.} \ \sigma \in \Gamma_-
  \big\}
  \,,
\end{aligned}
\end{equation}
where $u(\sigma)$ with~$\sigma\in\Gamma_+\cup\Gamma_-$ is understood
as the trace of the function~$u$
on that part of the boundary~$\partial\Omega$ (\cf~Remark~\ref{Rem.trace} below).
The associated sesquilinear form is symmetric, densely defined, closed
and bounded from below (the latter is not obvious unless $\tilde{\alpha} \geq 0$,
but it follows from the results~(\ref{inter.bound}) and~(\ref{lower.bound}) below).
Consequently, $Q_{\kappa,\alpha}$ gives rise (\cf~\cite[Sec.~VI.2]{Kato})
to a unique self-adjoint bounded-from-below operator
which we denote by~$-\Delta_{\kappa,\alpha}$.
It can be verified that~$-\Delta_{\kappa,\alpha}$ acts as
the classical Laplacian with the boundary conditions described
in Section~\ref{Sec.Intro} provided~$\Gamma$ is sufficiently regular.

It follows from assumptions~(\ref{Ass.Basic}) that
$\tubemapalt:I\times(-a,a)\to\Omega:\{(s,t)\mapsto\tubemap(s,t)\}$
is a $\Smooth^1$-diffeomorphism.
Consequently, $\Omega$~can be identified with the Riemannian manifold
$I \times (-a,a)$ equipped with the metric
$
  G_{ij} := (\partial_i\tubemap)\cdot(\partial_j\tubemap)
$,
where $i,j\in\{1,2\}$
and the \emph{dot} denotes the scalar product in~$\Real^2$.
Employing the Frenet formula
$
  \dot{N} = -\kappa \, \dot{\Gamma}
$,
one easily finds that
$
  (G_{ij}) = \diag(g_\kappa^2,1)
$,
where
\begin{equation}\label{Jacobian}
  g_\kappa(s,t) := 1-\kappa(s)\,t
\end{equation}
is the Jacobian of~$\tubemapalt$.

It follows that
$
  g_\kappa(s,t) \, ds \, dt
$
is the area element of the strip,
$\sii(\Omega)$ can be identified
with the Hilbert space
\begin{equation}\label{Hilbert}
  \sii\big(
  I \times (-a,a),g_\kappa(s,t) \, ds \, dt
  \big)
\end{equation}
and $-\Delta_{\kappa,\alpha}$ is unitarily equivalent
to the operator~$H_{\kappa,\alpha}$ on~(\ref{Hilbert})
associated with the quadratic form
\begin{equation}\label{form.h}
\begin{aligned}
  h_{\kappa,\alpha}[\psi]
  &:= \big\|g_\kappa^{-1}\partial_1\psi\big\|_\kappa^2
  + \big\|\partial_2\psi\big\|_\kappa^2
  + \int_{\Real} \alpha(s) \, |\psi(s,a)|^2 \, g_\kappa(s,a) \, ds
  \,,
  \\
  \psi\in\Dom(h_{\kappa,\alpha})
  &:= \big\{
  \psi\in\Sobi\big(\Real\times(-a,a)\big)
  \big|\
  \psi(s,-a) = 0 \quad \mbox{for a.e. } s\in\Real
  \big\}
  \,.
\end{aligned}
\end{equation}
Here $\|\cdot\|_{\kappa}$ stands for the norm in~(\ref{Hilbert})
and $\psi(s,\pm a)$ means the trace of the function~$\psi$
on the part of the boundary $I\times\{\pm a\}$
(\cf~Remark~\ref{Rem.trace} below).
In fact,
if the curve~$\Gamma$ is sufficiently smooth,
then~$H_{\kappa,\alpha}$ acts as the Laplace-Beltrami operator
$
  -G^{-1/2} \partial_i G^{1/2} G^{ij} \partial_j
$,
where $(G^{ij})=(G_{ij})^{-1}$ and $G:=\det(G_{ij})$,
but we will not use this fact.
Finally, let us notice
that the first assumption of~(\ref{Ass.Basic}) yields
$$
  0 < 1-\|\kappa\|_\infty\,a
  \leq g_\kappa(s,t) \leq
  1+\|\kappa\|_\infty\,a < 2
$$
uniformly in $(s,t) \in I\times(-a,a)$,
and that is actually why we can indeed write
$\Sobi\big(I \times (-a,a)\big)$ instead of
$
  \Sobi\big(
  I \times (-a,a),
  g_\kappa(s,t) \, ds \, dt
  \big)
$
in~(\ref{form.h}).

\begin{Remark}\label{Rem.trace}
The traces of $\psi\in\Sobi\big(I \times (-a,a)\big)$
on the boundary of the strip $I \times (-a,a)$
are well defined and square integrable (\cf~\cite{Adams}).
In particular, the boundary integral appearing in~(\ref{form.h}) is finite
(recall that~$\alpha$ is assumed to be bounded).
To ensure that the traces and the boundary integral
appearing in~(\ref{form.Laplacian}) are well defined too,
it is sufficient to notice that one can construct
traces of $u\in\Sobi(\Omega)$ to $\Gamma_+\cup\Gamma_-$
by means of the diffeomorphism~$\tubemapalt$,
the trace operator for the straight strip $I \times (-a,a)$
and inverses of the boundary mappings $\tubemap(\cdot,\{\pm a\})$.
The latter exists due to the second hypothesis in~(\ref{Ass.Basic}),
which is in fact a bit stronger than an analogous assumption
in the uniform Dirichlet case~\cite{DE,EFK}
(there it is enough to assume that
$\tubemap\upharpoonright I\times(-a,a)$ is injective).
In this context, one should point out that the approach
used by Daners in~\cite{Daners_2000}
makes it possible to deal with
Robin boundary conditions with positive~$\alpha$
on arbitrary bounded domains,
without using traces.
\end{Remark}
%

\section{An intermediate lower bound}\label{Sec.Inter}
%
In this section, we derive the central
lower bound of the present paper,
\ie~inequality (\ref{inter.bound}) below,
and explain its connection with Definition~\ref{Def.lambda}.

Neglecting in~(\ref{form.h})
the ``longitudinal kinetic energy'',
\ie~the term $\|g_\kappa^{-1}\partial_1\psi\|_\kappa$
in the expression for $h_{\kappa,\alpha}[\psi]$,
and using Fubini's theorem,
one immediately gets
\begin{equation}\label{inter.bound}
  \inf\sigma(H_{\kappa,\alpha})
  \geq \inf_{s \in I}\lambda\big(\kappa(s),\alpha(s)\big)
  \,,
\end{equation}
where $\lambda(\kappa,\alpha)$
denotes the first eigenvalue of the self-adjoint
one-dimensional operator~$B_{\kappa,\alpha}$ on
$$
  \Hilbert_\kappa:=\sii\big((-a,a),(1-\kappa\,t) dt\big)
$$
associated with the quadratic form
\begin{equation}\label{form.b}
\begin{aligned}
  b_{\kappa,\alpha}[\psi] &:= \int_{-a}^a |\psi'(t)|^2 \, (1-\kappa\,t) dt
  + \alpha \, |\psi(a)|^2 \, (1-\kappa\,a)
  \,,
  \\
  \psi \in \Dom(b_{\kappa,\alpha}) &:=
  \big\{
  \psi \in \Sobi((-a,a)) \, \big| \
  \psi(-a)=0
  \big\}
  \,.
\end{aligned}
\end{equation}
With a slight abuse of notation,
we denote by $\kappa\in(-1/a,1/a)$
and~$\alpha\in\Real$ given constants now.
One easily verifies that
\begin{equation}\label{op.B}
\begin{aligned}
  (B_{\kappa,\alpha}\psi)(t) &= -\psi''(t) + \frac{\kappa}{1-\kappa\,t}\,\psi'(t)
  \,,
  \\
  \psi \in \Dom(B_{\kappa,\alpha}) &=
  \big\{
  \psi \in \sobi((-a,a)) \, \big| \
  \psi(-a)=0 \ \, \& \ \, \psi'(a)+\alpha\,\psi(a)=0
  \big\}
  \,.
\end{aligned}
\end{equation}
Note that the values of~$\psi$ and~$\psi'$
at the boundary points of~$(-a,a)$
are well defined due to the Sobolev embedding theorem.

$B_{\kappa,\alpha}$ is clearly a positive operator for~$\alpha \geq 0$.
Furthermore, using the elementary inequality
$
  |\psi(a)|^2
  \leq \varepsilon \int_{-a}^a |\psi'(t)|^2 dt
  + \varepsilon^{-1} \int_{-a}^a |\psi(t)|^2 dt
$
with~$\varepsilon>0$, it can be easily shown that
\begin{equation}\label{lower.bound}
  \lambda(\kappa,\alpha)
  \geq - \alpha^2 \, \frac{(1+|\kappa|\,a)^2}{(1-|\kappa|\,a)^2}
  \,,
\end{equation}
\ie, $B_{\kappa,\alpha}$ is bounded from below in any case.
This and~(\ref{inter.bound}) prove
that~$H_{\kappa,\alpha}$ (and therefore~$-\Delta_{\kappa,\alpha}$)
is bounded from below \emph{a fortiori}.

Using coordinates analogous to~(\ref{StripMap})
and the circular (respectively straight) symmetry,
it is easy to see that~$B_{\kappa,\alpha}$ is nothing else
than the ``radial" (respectively ``transversal'') part
of the Laplacian on $\sii(\Annulus_\kappa)$ if $\kappa\not=0$
(respectively $\kappa=0$) in Definition~\ref{Def.lambda}.
(We refer to~\cite[Lemma~4.1]{EFK} for more details on the partial
wave decomposition in the case~$\alpha=+\infty$.)
This shows that the geometric Definition~\ref{Def.lambda} of~$\lambda$
and the definition via~(\ref{form.b}) are in fact equivalent.

In view of~(\ref{inter.bound}), it remains to establish
the monotonicity properties of~$\lambda$
stated in Theorem~\ref{thm.annulus}
in order to prove Theorem~\ref{Thm.bound}.
This will be done in the next section.

\section{Dirichlet-Robin annuli}\label{Sec.Annulus}
%
Using standard arguments (\cf~\cite[Sec.~8.12]{Gilbarg-Trudinger}),
one easily shows that~$\lambda(\kappa,\alpha)$,
as the lowest eigenvalue of~$B_{\kappa,\alpha}$,
is simple and has a positive eigenfunction.
We denote the latter by $\psi_{\kappa,\alpha}$
and normalize it to have unit norm in
the Hilbert space~$\Hilbert_\kappa$.

\subsection{Dependence on~$\alpha$}\label{Sec.monotonicity.alpha}
%
The first property of Theorem~\ref{thm.annulus}
follows directly from the variational definition
of $\lambda(\kappa,\alpha)$.
In detail, using $\psi_{\kappa,\alpha+\delta}$ with any $\delta>0$
as a test function for $\lambda(\kappa,\alpha)$,
we get
\begin{equation}\label{monotonicity.alpha}
  \lambda(\kappa,\alpha) \leq \lambda(\kappa,\alpha+\delta)
  - \delta \, \psi_{\kappa,\alpha+\delta}(a)^2 \, (1-\kappa\,a)
  < \lambda(\kappa,\alpha+\delta)
  \,,
\end{equation}
\ie\ $\alpha\mapsto\lambda(\kappa,\alpha)$ is increasing.
Note that the \emph{strict} monotonicity is a consequence of the fact
that $\psi_{\kappa,\alpha+\delta}\in\Dom(B_{\kappa,{\alpha+\delta}})$;
indeed, $\psi_{\kappa,\alpha+\delta}(a)=0$ would imply that
$\psi_{\kappa,\alpha+\delta}'(a)=0$  also,
giving a contradiction.
Using now $\psi_{\kappa,\alpha}$
as a test function for $\lambda(\kappa,\alpha+\delta)$,
we get
\begin{equation}
  \lambda(\kappa,\alpha+\delta) \leq \lambda(\kappa,\alpha)
  + \delta \, \psi_{\kappa,\alpha}(a)^2 \, (1-\kappa\,a)
  \xrightarrow[\delta \to 0]{} \lambda(\kappa,\alpha)
  \,,
\end{equation}
which, together with~(\ref{monotonicity.alpha}),
gives the continuity of $\lambda$ in the second variable.

\subsection{Dependence on~$\kappa$}
%
Not all of the other properties of Theorem~\ref{thm.annulus}
are so obvious from the variational definition of $\lambda(\kappa,\alpha)$
via~$B_{\kappa,\alpha}$ because the Hilbert space~$\Hilbert_\kappa$
depends on~$\kappa$.
To overcome this, we introduce the unitary transformation
\begin{equation}\label{unitary}
  U_\kappa: \Hilbert_\kappa \to \Hilbert_0:  \
  \big\{\psi \mapsto (1-\kappa\,t)^\demi\psi\big\}
\end{equation}
and the unitarily equivalent operator
$
  \hat{B}_{\kappa,\alpha}:=U_\kappa B_{\kappa,\alpha}U_\kappa^{-1}
$
associated with the transformed form
$
  \hat{b}_{\kappa,\alpha}[\cdot] := b_{\kappa,\alpha}[U_\kappa^{-1}\cdot]
$.
Given any
$
  \phi\in\Dom(\hat{b}_{\kappa,\alpha})
$,
we insert $\psi = U_\kappa^{-1} \phi$ into~(\ref{form.b}),
integrate by parts and finds
\begin{equation}\label{form.Ub}
  \hat{b}_{\kappa,\alpha}[\phi]
  = \int_{-a}^a |\phi'(t)|^2 dt
  - \int_{-a}^a \frac{\kappa^2}{4(1-\kappa\,t)^2} \, |\phi(t)|^2 dt
  + \left(\alpha+\frac{\kappa}{2(1-\kappa\,a)}\right) |\phi(a)|^2
  \,.
\end{equation}
We also verify that
\begin{equation}\label{op.UB}
\begin{aligned}
  (\hat{B}_{\kappa,\alpha}\phi)(t)
  &= -\phi''(t) - \frac{\kappa^2}{4(1-\kappa\,t)^2}\,\phi(t)
  \,,
  \\
  \phi \in \Dom(\hat{B}_{\kappa,\alpha}) &=
  \Big\{
  \phi \in \sobi((-a,a)) \, \big| \
  \phi(-a)=0 \ \,
  \\
  & \rule{15ex}{0ex} \& \ \,
  \phi'(a)+\left(\alpha+\mbox{$\frac{\kappa}{2(1-\kappa\,a)}$}\right) \phi(a)=0
  \Big\}
  \,.
\end{aligned}
\end{equation}
It is important to notice that while $\Dom(B_{\kappa,\alpha})$
is not invariant under~$U_\kappa$,
one still has $\Dom(\hat{b}_{\kappa,\alpha})=\Dom(b_{\kappa,\alpha})$.

\subsubsection{Continuity}\label{Sec.continuity}
%
Following~\cite[Sec.~VII.~4]{Kato},
$\kappa\mapsto\hat{b}_{\kappa,\alpha}$ forms a holomorphic
family of forms of type~(a)
and $\kappa\mapsto\hat{B}_{\kappa,\alpha}$ forms
a self-adjoint holomorphic family of operators of type~(B).
In particular, $\kappa\mapsto\lambda(\kappa,\alpha)$ is continous,
which proves~(ii) of Theorem~\ref{thm.annulus}.
Moreover, denoting by $\phi_{\kappa,\alpha}:=U_\kappa\psi_{\kappa,\alpha}$
the eigenfunction of $\hat{B}_{\kappa,\alpha}$
corresponding to $\lambda(\kappa,\alpha)$,
we get that $\kappa\mapsto\phi_{\kappa,\alpha}$ is continous
in the norm of~$\Hilbert_0$.
In view of~(\ref{unitary}), it then follows that also
$\kappa\mapsto\psi_{\kappa,\alpha}$ is continuous in the norm of~$\Hilbert_0$.

\subsubsection{Monotonicity}
%
Since the function
$
  f:\kappa \mapsto \frac{\kappa}{1-\kappa\,t}
$
is increasing on $(-1/a,1/a)$ for any $t\in[-a,a]$,
one easily verifies Theorem~\ref{thm.annulus}.(iii)
by means of the variational definition of $\lambda(\kappa,\alpha)$
via~$\hat{B}_{\kappa,\alpha}$ and an argument similar to that
used in Section~\ref{Sec.monotonicity.alpha}.

However, the above argument fails to prove~(iv) of Theorem~\ref{thm.annulus}
because $-f^2$ is decreasing on $[0,1/a)$,
so that one gets an interplay between the increasing boundary term
and decreasing potential in~(\ref{form.Ub}) for positive curvatures.
Therefore we come back to the initial operator~(\ref{op.B})
and calculate the derivative of $\kappa\mapsto\lambda(\kappa,\alpha)$:
\begin{Lemma}\label{Lem.derivative}
$\forall \kappa \in (-1/a,1/a)$, \ $\forall\alpha\in\Real$, \
\begin{equation}\label{derivative}
  \frac{\partial\lambda}{\partial\kappa}\,(\kappa,\alpha)
  = \int_{-a}^a \frac{\psi_{\kappa,\alpha}(t)\,\psi_{\kappa,\alpha}'(t)}{1-\kappa\,t}
  \, dt
  \,.
\end{equation}
\end{Lemma}
\begin{proof}
Throughout this proof, we omit the dependence
of~$\lambda$ and the corresponding eigenfunction on~$\alpha$.

We write the eigenvalue equation for~$B_{\kappa,\alpha}$
with~$\psi_\kappa$ and~$\lambda(\kappa)$ as
\begin{equation}\label{ev.eq}
  -\big[\psi_\kappa'(t) \, (1-\kappa\,t)\big]'
  = \lambda(\kappa) \, \psi_\kappa(t) \, (1-\kappa\,t)
\end{equation}
and consider the analogous equation at~$\kappa+\delta$,
with $\delta\in\Real\setminus\{0\}$
so small that $|\kappa+\delta|\,a$ is less than~$1$.
Multiplying~(\ref{ev.eq}) by $\psi_{\kappa+\delta}$,
integrating by parts, combining the result
with the result coming from analogous manipulations
applied to the problem at~$\kappa+\delta$,
dividing by~$\delta$,
integrating by parts once more
and using the eigenvalue equation for~$B_{\kappa,\alpha}$,
we arrive at
\begin{eqnarray*}
  \lefteqn{\frac{\lambda(\kappa+\delta)-\lambda(\kappa)}{\delta}
  \int_{-a}^a \psi_\kappa(t)\,\psi_{\kappa+\delta}(t)\,(1-\kappa\,t)\,dt}
  \\
  & = & \lambda(\kappa+\delta)
  \int_{-a}^a \psi_\kappa(t)\,\psi_{\kappa+\delta}(t)\,t\,dt
  \\
  &&
  - \int_{-a}^a \psi_\kappa'(t)\,\psi_{\kappa+\delta}'(t)\,t\,dt
  - \alpha\,a\,\psi_\kappa(a)\,\psi_{\kappa+\delta}(a)
  \\
  & = & \big[\lambda(\kappa+\delta)-\lambda(\kappa)\big]
  \int_{-a}^a \psi_\kappa(t)\,\psi_{\kappa+\delta}(t)\,t\,dt
  + \int_{-a}^a \frac{\psi_\kappa'(t)\,\psi_{\kappa+\delta}(t)}
  {1-\kappa\,t} \, dt
  \,.
\end{eqnarray*}
Letting~$\delta$ go to zero yields the desired result
by means of the continuity of $\kappa\mapsto\lambda(\kappa)$
and $\kappa\mapsto\psi_\kappa$ established in Section~\ref{Sec.continuity}.
\end{proof}

Lemma~\ref{Lem.derivative} yields~(iv) of Theorem~\ref{thm.annulus}
whenever the integral on the right hand side of~(\ref{derivative})
is positive. In particular, this is the case
when~$\psi_{\kappa,\alpha}'$ is non-negative:
\begin{Lemma}
$\forall\kappa \in (-1/a,1/a)$, \
$\forall \alpha \in (-\infty,0]$, \
$$
  t\mapsto\psi_{\kappa,\alpha}(t) : (-a,a) \to \Real
  \quad \mbox{is increasing}.
$$
\end{Lemma}
\begin{proof}
Throughout this proof,
we omit the dependence of~$\lambda$
and the corresponding eigenfunction
on~$\kappa$ and~$\alpha$.

Since~$\psi$ is a positive eigenfunction and $\psi(-a)=0$,
respectively $\psi'(a)=-\alpha\psi(a)$,
we know that $\psi'(-a)>0$, respectively $\psi'(a) \geq 0$.
Recall also that $\psi(a)>0$.
We claim that $\psi'>0$ on~$(-a,a)$.

\emph{Case} $\lambda<0$.
The eigenvalue problem for~(\ref{op.B})
implies that if $\psi'(t)=0$ for some $t\in(-a,a)$,
then~$\psi''(t)>0$, \ie\ $\psi$ has a local minimum at~$t$.
Consequently, if there exists a~$t_1\in(-a,a)$ such that $\psi'(t_1)=0$,
then, since $\psi'(-a)>0$,
there must also be a $t_2\in(-a,t_1)$ such that
$\psi$ has a local maximum at~$t_2$, a contradiction.

\emph{Case} $\lambda>0$.
The eigenvalue problem for~(\ref{op.B})
implies that if $\psi'(t)=0$ for some $t\in(-a,a]$,
then~$\psi''(t)<0$, \ie~$\psi$ has a local maximum at~$t$.
Consequently, if there exists a~$t_1\in(-a,a)$ such that $\psi'(t_1)=0$,
then, since $\psi'(a) \geq 0$,
there must also be a $t_2\in(t_1,a]$ such that
$\psi'(t_2)=0$ and $\psi'<0$ on $(t_1,t_2)$,
\ie\ $\psi$ does not have a local maximum at~$t_2$,
a contradiction.

\emph{Case} $\lambda=0$.
Integrating~(\ref{ev.eq}), we get
$
  \psi'(t) = -\alpha\,\frac{1-\kappa\,a}{1-\kappa\,t} \, \psi(a) > 0
$
for all $t\in[-a,a]$ (the equality would imply a trivial eigenfunction).
\end{proof}
%

\subsubsection{Boundary values}
%
Using the geometrical meaning of $\lambda(\kappa,\alpha)$
(\cf~Definition~\ref{Def.lambda}) and since~$\Annulus_\kappa$
converges (\eg, in the sense of metrical convergence~\cite{RT})
to the disc~$\Ball(2a)$ with the central point removed as $|\kappa| \to 1/a$,
the limits in Theorem~\ref{thm.annulus}.(v)
are natural to expect. We prove each of them separately.

\paragraph{The negative limit}
The limit value for $\lambda(\kappa,\alpha)$
as \mbox{$\varepsilon:=-(\kappa^{-1}+a) \to 0$}
follows from Flucher's paper~\cite{Flucher},
where an approximation formula for eigenvalues
in domains with spherical holes is found.
The only difference is the fact that in our case the boundary
of the domain also changes as $\eps$ goes to zero.
We overcome this complication by transforming the eigenvalue problem
for the Laplacian on~$\Annulus_\kappa$ into
\begin{equation}\label{ev1}
\left\{
\begin{aligned}
  -\Delta u &= \lambda_\varepsilon(\alpha_\varepsilon) u
  &\mbox{in}\qquad& \Annulus\big(\varepsilon(2a+\varepsilon)^{-1},1\big) \,,
  \\
  u &= 0
  &\mbox{on}\qquad& \partial\Ball\big(\varepsilon(2a+\varepsilon)^{-1}\big) \,,
  \\
  \frac{\partial u}{\partial N} + \alpha_\varepsilon \,u &= 0
  &\mbox{on}\qquad& \partial\Ball(1) \,,
\end{aligned}
\right.
\end{equation}
where
$
  \lambda_\varepsilon(\alpha_\varepsilon)
  := (2a+\varepsilon)^2\lambda(-(a+\varepsilon)^{-1},\alpha)
$,
$
  \alpha_\varepsilon
  := (2a+\varepsilon)\alpha
$
and $N$ is the outward unit normal on~$\partial\Ball(1)$.
By the minimax principle,
$$
  \lambda_\varepsilon\big(\alpha_{-(\sgn\alpha)\varepsilon_0}\big)
  \leq \lambda_\varepsilon(\alpha_\varepsilon) \leq
  \lambda_\varepsilon\big(\alpha_{(\sgn\alpha)\varepsilon_0}\big)
$$
for any fixed $\varepsilon_0\in(\varepsilon,2a)$,
where~$\lambda_\varepsilon(\alpha_{\pm\varepsilon_0})$
denotes the eigenvalue of the problem~(\ref{ev1}) with~$\alpha_\varepsilon$
being replaced by~$\alpha_{\pm\varepsilon_0}$.
Then it is clear that
$
  \lambda_\varepsilon(\alpha_\varepsilon)
  \to (2a)^2 \nu(\alpha)
$
as $\varepsilon \to 0$ because
it is true for~$\lambda_\varepsilon(\alpha_{\pm\varepsilon_0})$
by~\cite{Flucher} and~$\varepsilon_0$ can be chosen arbitrarily small.

\paragraph{The positive limit}
If~$\alpha>0$, the limit value for $\lambda(\kappa,\alpha)$
as \mbox{$\kappa^{-1} \to a$} could be derived
by means of a paper by Dancer and Daners, \cite{Dancer-Daners_1997},
where they study domain perturbations
for elliptic equations subject to Robin boundary condtions.
However, since they restrict to positive~$\alpha$ and we do not know
about a similar perturbation result for~$\alpha<0$,
we establish the limit value by rather elementary considerations.

Assuming~$\kappa\not=0$,
the eigenvalue problem for~$B_{\kappa,\alpha}$ is explicitly solvable
in terms of the Bessel functions~$J_0$ and~$Y_0$
(\cf~\cite[Chap.~7]{Wang-Guo})
and the eigenvalue~$\lambda(\kappa,\alpha)$ is then determined as
the smallest (in absolute value) zero~$\lambda$ of the implicit equation
\begin{multline}\label{implicit}
  J_0\big(\sqrt{\lambda} (1 + \kappa a)/\kappa\big)
  \left[
  \sqrt{\lambda} Y_1\big(\sqrt{\lambda} (1 - \kappa a)/\kappa\big)
  + \alpha Y_0\big(\sqrt{\lambda} (1 - \kappa a)/\kappa\big)
  \right]
  \\
  = Y_0\big(\sqrt{\lambda} (1 + \kappa a)/\kappa\big)
  \left[
  \sqrt{\lambda} J_1\big(\sqrt{\lambda} (1 - \kappa a)/\kappa\big)
  + \alpha J_0\big(\sqrt{\lambda} (1 - \kappa a)/\kappa\big)
  \right]
  \,.
\end{multline}
Although the case~$\lambda(\kappa,\alpha)=0$ should be treated separately,
a formal asymptotic expansion of~(\ref{implicit}) around~$\sqrt{\lambda}=0$
also gives the correct condition for a zero eigenvalue:
\begin{equation}
  \lambda(\kappa,\alpha) = 0
  \quad\Longleftrightarrow\quad
  \kappa = \alpha \, (1-\kappa\,a) \log\frac{1-\kappa\,a}{1+\kappa\,a}
  \,.
\end{equation}
In particular, the condition yields
that for any $\alpha<-1/(2a)$ there always exists~$\kappa_0\in(0,1/a)$
such that $\lambda(\kappa_0,\alpha)=0$.
This and the properties~(i), (ii) and~(iv) of Theorem~\ref{Thm.bound}
imply that
$
  \lim_{\kappa \to 1/a} \lambda(\kappa,\alpha) > 0
$
for any~$\alpha\in\Real$.
We also know that the limit is bounded because
$
  \lambda(\kappa,\alpha) < \lambda(\kappa,+\infty)
$
by the minimax principle and $\lambda(\kappa,+\infty)$ tends
to the first eigenvalue of the Dirichlet Laplacian
in the disc $\Ball(2a)$, \ie~$\nu(+\infty) \equiv j_{0,1}^2/(2a)^2$,
as $\kappa \to 1/a$
by known convergence theorems (\cf~one of~\cite{RT,Stollmann,Daners}).
Applying the limit to~(\ref{implicit}),
we get a bounded value on the right hand side,
while the left hand side admits the asymptotic expansion
$
  -\frac{\sqrt{\lambda}}{\pi} \, J_0(\sqrt{\lambda}\,2a) \,
  \big[\sqrt{\lambda}\,(1-\kappa\,a)/(2\kappa)\big]^{-1}
  +\mathcal{O}(\kappa^{-1}-a)
$.
That is, $\sqrt{\lambda}\,2a$ necessarily converges
to the first zero of the Bessel function~$J_0$ as $\kappa \to 1/a$.

\section{Infinite strips}\label{Sec.Infinite}
%
Let $I=\Real$ throughout this section.
The proof of Theorem~\ref{Thm.Hardy}
is based on the following two lemmata.

Firstly, Theorem~\ref{thm.annulus} implies:
\begin{Lemma}\label{Lem.crucial}
Assume the hypotheses of Theorem~\ref{Thm.Hardy}.
Then the function~$\mu:\Real\to\Real$
defined by
$$
  s \,\mapsto\, \mu(s)
  := \lambda\big(\kappa(s),\alpha(s)\big)
  - \lambda(0,\alpha_0)
$$
is continuous, non-zero and non-negative.
\end{Lemma}
\noindent
Hereafter we shall use the same notation~$\mu$
for the function $\mu \otimes 1$ on $\Real\times(-a,a)$.

Secondly, we shall need the following
Hardy-type inequality for a Schr\"odinger operator in a strip
with the potential being a characteristic function:
\begin{Lemma}\label{Lem.Hardy}
For any $\psi\in\Sobi\big(\Real\times(-a,a)\big)$,
\begin{equation*}
  \int_{\Real\times(-a,a)}
  \rho^{-2} \, |\psi|^2
  \ \leq \
  16 \int_{\Real\times(-a,a)} |\partial_1\psi|^2
  + \big(2+64/|J|^2\big) \int_{J\times(-a,a)} |\psi|^2
  \,,
\end{equation*}
where
$
  \rho(s,t) := \sqrt{1+(s-s_0)^2}
$,
$J$~is any bounded subinterval of~$\Real$
and~$s_0$ is the mid-point of~$J$.
\end{Lemma}
\noindent
This lemma can be established quite easily
by means of the classical one-dimensional Hardy inequality
$
  \int_{\Real} x^{-2} |v(x)|^2 \, dx
  \leq 4 \int_{\Real}\,|v'(x)|^2\, dx
$
valid for any $v\in\Sobi(\Real)$ with $v(0)=0$
and Fubini's theorem;
we refer the reader to \cite[Sec.~3.3]{EKK}
or \cite[proof of Lem.~2]{K3}
for more details.

\subsection{Proof of Theorem~\ref{Thm.Hardy}}\label{Sec.Proof1}
%
Let~$\psi$ belong to the dense subspace of $\Dom(h_{\kappa,\alpha})$
given by $\Smooth^\infty$-smooth functions
on $\Real\times(-a,a)$ which vanish in a neighbourhood
of $\Real\times\{-a\}$ and which are restrictions of
functions from $\Smooth_0^\infty(\Real^2)$.
Assume the hypotheses of Theorem~\ref{Thm.Hardy}
so that the conclusions of Lemma~\ref{Lem.crucial} hold.
Let~$J$ be any closed subinterval of~$\Real$
on which~$\mu$ defined in Lemma~\ref{Lem.crucial} is positive.

The first step is to come back to the intermediate
lower bound~(\ref{inter.bound});
we also use the definition of~$\lambda$ via~(\ref{form.b}),
but we do not neglect the ``longitudinal kinetic energy'':
\begin{align*}
  h_{\kappa,\alpha}[\psi]
  - \lambda(0,\alpha_0) \, \|\psi\|_{\kappa}^2
  & \geq
  \big\|g_\kappa^{-1}\partial_1\psi\big\|_{\kappa}^2
  + \big\|\mu^{1/2}\psi\big\|_{\kappa}^2
  \\
  & \geq
  \big\|g_\kappa^{-1}\partial_1\psi\big\|_{\kappa}^2
  + \epsilon \, (1-\|\kappa\|_\infty a) \, \min_J \mu
  \int_{J\times(-a,a)} \!\! |\psi|^2
  \,.
\end{align*}
Here $\epsilon\in(0,1]$ is arbitrary for the time being.
Applying Lemma~\ref{Lem.Hardy} to the last integral,
we arrive at
\begin{eqnarray*}
  \lefteqn{h_{\kappa,\alpha}[\psi]
  - \lambda(0,\alpha_0) \, \|\psi\|_{\kappa}^2}
  \\
  && \geq
  \left(
  \frac{1}{1+\|\kappa\|_\infty a}
  - \frac{16 \, \epsilon \, (1-\|\kappa\|_\infty a) \, \min_J \mu}
  {2+64/|J|^2}
  \right) \,
  \int_{\Real\times(-a,a)} |\partial_1\psi|^2
  \\
  && \phantom{\geq}
  + \frac{\epsilon \, (1-\|\kappa\|_\infty a) \, \min_J \mu}
  {2+64/|J|^2} \,
  \int_{\Real\times(-a,a)} \rho^{-2} \, |\psi|^2
  \,.
\end{eqnarray*}
Choosing now~$\epsilon$ as the minimum between~$1$
and the value such that the first term on the right hand side
of the last estimate vanishes,
we finally get
\begin{equation}\label{Hardy}
  h_{\kappa,\alpha}[\psi]
  - \lambda(0,\alpha_0) \, \|\psi\|_{\kappa}^2
  \ \geq \
  c \, \big\|\rho^{-1}\psi\big\|_{\kappa}^2
\end{equation}
with
\begin{equation}\label{constant}
  c \ := \
  \min
  \left\{
  \frac{(1-\|\kappa\|_\infty a) \, \min_J \mu}
  {\big(2+64/|J|^2\big) \, (1+\|\kappa\|_\infty a)}
  \, , \,
  \frac{1}{16 \, (1+\|\kappa\|_\infty a)^2}
  \right\}
  \,.
\end{equation}
In view of Section~\ref{Sec.Laplacian},
we conclude that~(\ref{Hardy})~is equivalent to
\begin{equation}\label{Hardy.bound.exact}
  Q_{\kappa,\alpha}[u]
  - \lambda(0,\alpha_0) \, \|u\|_{\sii(\Omega)}^2
  \ \geq \
  c \, \big\|(\rho\circ\tubemap)^{-1} u \big\|_{\sii(\Omega)}^2
\end{equation}
for all $u\in\Dom(Q_{\kappa,\alpha})$,
which is the exact meaning of~(\ref{Hardy.bound}).

\subsection{Proof of Corollary~\ref{Thm.stability}}\label{Sec.Proof2}
%
Let~$\psi$ be as in the previous section.
The present proof is based on an algebraic comparison of
$h_{\kappa,0}[\psi]-\lambda(0,0)\;\!\|\psi\|_{\kappa}^2$
with $h_{\kappa_+,0}[\psi]-\lambda(0,0)\;\!\|\psi\|_{\kappa_+}^2$
and a usage of~(\ref{Hardy}).

For every $(s,t)\in\Real\times(-a,a)$,
we have
$$
  1 - f_\eps(s)
  \ \leq \
  \frac{g_\kappa(s,t)}{g_{\kappa_+}(s,t)}
  \ \leq \
  1 + f_\eps(s)
  \qquad\mbox{with}\qquad
  f_\eps(s) := \frac{\eps\,a\,\chi_I(s)}{1-\|\kappa_+\|_\infty a}
  \,,
$$
where~$\chi_I$ denotes the characteristic function
of the set $I\times(-a,a)$.
Hereafter we assume $\eps \leq (1-\|\kappa_+\|_\infty a)/(2a)$
so that the lower bound is greater or equal to~$1/2$.
Using the same notation~$f_\eps$
for the functions $f_\eps \otimes 1$ on $\Real\times(-a,a)$,
we have
\begin{eqnarray*}
  \lefteqn{
  h_{\kappa,0}[\psi] - \lambda(0,0)\;\!\|\psi\|_\kappa^2
  }
  \\
  && \ \geq \
  \int_{\Real\times(-a,a)}
  (1+f_\eps)^{-1} \, g_{\kappa_+}^{-1} \, |\partial_1\psi|^2
  \\
  && \phantom{\ \geq \ }
  + \int_\Real ds \, \big(1-f_\eps(s)\big)
  \int_{-a}^a dt \, g_{\kappa_+}(s,t) \left(
  |\partial_2\psi(s,t)|^2 - \lambda(0,0)\;\! |\psi(s,t)|^2
  \right)
  \\
  && \phantom{\ \geq \ }
  - \lambda(0,0) \int_{\Real\times(-a,a)}
  2 f_\eps \, g_{\kappa_+} \, |\psi|^2
  \,.
\end{eqnarray*}
Recalling the definition of~$\lambda$
via~(\ref{form.b}) and Lemma~\ref{Lem.crucial},
it is clear that the term in the second line
after the inequality sign is non-negative.
Consequently,
\begin{align*}
  h_{\kappa,0}[\psi]
  - \lambda(0,0)\;\!\|\psi\|_\kappa^2
  & \,\geq\,
  \frac{1}{2}
  \left(
  h_{\kappa_+,0}[\psi]
  - \lambda(0,0)\;\!\|\psi\|_{\kappa_+}^2
  \right)
  \\
  & \phantom{\,\geq\,}
  - \, \lambda(0,0) \int_{\Real\times(-a,a)}
  2 f_\eps \, g_{\kappa_+} \, |\psi|^2
  \,.
\end{align*}
Using~(\ref{Hardy}) with~$\alpha$ being equal to~$0$,
with~$\kappa$ being replaced by~$\kappa_+$
and with~$s_0$ being from the support of~$\kappa_+$,
we finally obtain
\begin{equation}\label{Hardy.c}
  h_{\kappa,0}[\psi]
  - \lambda(0,0)\;\!\|\psi\|_\kappa^2
  \ \geq \
  \big\|w^{1/2}\psi\big\|_{\kappa}^2
  \,,
\end{equation}
where
\begin{equation*}
  w(s,t) :=
  \frac{c/4}{1+(s-s_0)^2}
  - \lambda(0,0) \,
  \frac{\eps\,a\,\chi_I(s)}{1-\|\kappa_+\|_\infty a}
\end{equation*}
is positive for all sufficiently small~$\eps$.
Equivalently,
\begin{equation}\label{Hardy.general}
  -\Delta_{\kappa,0}
  \ \geq \
  \lambda(0,0) + w\circ\mathcal{L}^{-1}
\end{equation}
in the sense of quadratic forms on~$\sii(\Omega)$.
This concludes the proof of Corollary~\ref{Thm.stability}.

\section{Remarks and open questions}\label{Sec.closing}
%
It follows immediately from the minimax principle that
the lower bound of Theorem~\ref{Thm.bound} also applies
to other boundary conditions imposed on~$L_\pm$,
\eg, Dirichlet, periodic, certain Robin, \etc.

Of course, it is also possible to impose Robin boundary conditions
on~$\Gamma_-$ instead of Dirichlet.
Then the lower bound of the type~(\ref{inter.bound}) still holds
and the problem is translated to the study of properties
of the first eigenvalue in a Robin-Robin annulus.
The techniques of the present paper will also apply
to certain values of the parameters in such a case. However,
we refrained from doing so to keep the statement of results
as simple as possible.

It follows from Theorem~\ref{thm.annulus} that~$\nu(\alpha)$
gives a uniform lower bound to the spectral threshold
of $-\Delta_{\kappa,\alpha}$ provided $\alpha \leq 0$ or $\kappa \leq 0$.
We conjecture this to be always the case,
but were not able to prove it in general.
In this context, it would be desirable to prove that
$\kappa\mapsto\lambda(\kappa,\alpha)$ does not possess local minima
for any~$\alpha\in\Real$.

We proved the fact that $\kappa\mapsto\lambda(\kappa,\alpha)$
is increasing on $(0,1/a)$ only for non-positive~$\alpha$.
It is clear from the limiting Dirichlet problem (\cf~\cite{EFK})
that this property will not hold for large positive~$\alpha$.
However, formula~(\ref{derivative}) suggests
that this is still true for small values of~$\alpha$.
Numerical results show (\cf~Figure~\ref{figure})
that the critical value is approximately~$0.78$ for $a=1$.

To transfer the numerical results of Figure~\ref{figure}
for different values of~$a$, it is sufficient to notice
that~$\lambda$ scales as:
$
  \lambda(\kappa,\alpha;a)
  = a^{-2} \lambda(\kappa a,\alpha a;1)
$.

Proposition~\ref{Prop.DK} contains just one example
of sufficient condition which guarantees
the existence of discrete eigenvalues
in infinite curved strips.
Further results can be obtained in the spirit of~\cite{DKriz2,KKriz}.
An open question is, \eg, whether the discrete spectrum exists
for certain strips with $\kappa>0$ and $\alpha>0$.
Let us recall that this is always the case for~$\alpha=+\infty$.

For simplicity, we assumed that~$\kappa$ and $\alpha-\alpha_0$
had compact support when we considered infinite strips.
However, the claim of Corollary~\ref{Corol.Infinite}
holds whenever the essential spectrum~(\ref{EssSpec}) is preserved,
and this might be checked under much less restrictive
conditions about the decay of~$\kappa$ and $\alpha-\alpha_0$
at infinity.
For instance, modifying the approach of~\cite{KKriz},
it should be enough just to require that the limits
at infinity are equal to zero.
In fact, Theorem~\ref{Thm.Hardy} holds without any condition
about the decay of~$\kappa$ and $\alpha-\alpha_0$ at infinity,
but it is of interest only in the case the essential spectrum
does not start above $\lambda(0,\alpha_0)$.
In any case, a fast decay of curvature at infinity
is needed to prove Corollary~\ref{Thm.stability};
namely, $\kappa(s)=\mathcal{O}(s^{-2})$ as $|s|\to\infty$.
This quadratic decay is related to the decay
of the Hardy weight in Theorem~\ref{Thm.Hardy},
which is typical for Hardy inequalities
involving the Laplacian,
and cannot be therefore improved by the present method.

Under suitable global geometric conditions
about the reference curve~$\Gamma$,
the intrinsic distance $|s-s_0|$
which appears in the function~$\rho$ of Theorem~\ref{Thm.Hardy}
can be estimated by an exterior one.
For instance, if~$\Gamma$ is an embedded unit-speed curve
with compactly supported curvature,
then it is easy to see that
there exists a positive number~$\delta$ such that
\begin{equation*}
  \forall s,s'\in\Real, \quad
  \delta \, |s-s'|
  \leq |\Gamma(s)-\Gamma(s')| \leq
  |s-s'|
  \,.
\end{equation*}

Corollary~\ref{Thm.stability} extends the class of strips
from~\cite{DKriz2} with empty discrete spectrum.
An open question is to decide whether an analogous
result holds for other~$\alpha$ satisfying
$\alpha_0 \leq \alpha \leq 0$.

\section*{Acknowledgements}
This work was partially supported by FCT (Portugal)
through projects POCI/\-MAT/60863/2004 (POCI2010)
and SFRH/BPD/11457/2002.
The second author (D.K.) was also supported by
the Czech Academy of Sciences and its Grant Agency
within the projects IRP AV0Z10480505 and A100480501,
and by the project LC06002 of the Ministry of Education,
Youth and Sports of the Czech Republic.

%
\addcontentsline{toc}{section}{References}
%
\providecommand{\bysame}{\leavevmode\hbox to3em{\hrulefill}\thinspace}
\providecommand{\MR}{\relax\ifhmode\unskip\space\fi MR }
\providecommand{\MRhref}[2]{%
  \href{http://www.ams.org/mathscinet-getitem?mr=#1}{#2}
}
\providecommand{\href}[2]{#2}

%
%
\addcontentsline{toc}{section}{List of Figures}
\listoffigures
\end{document}